\documentclass[pra,twocolumn,amsmath, superscriptaddress,notitlepage,showpacs]{revtex4-1}
\usepackage{amsmath,amsfonts,amssymb,amstext,amscd,amsthm,dsfont,bbm,hyperref}
\usepackage{physics}
\usepackage{color}
\usepackage{soul,xcolor}
\usepackage{gensymb}
\hypersetup{
    colorlinks = true,
    citecolor=blue,
    linkcolor = red,
    anchorcolor = red,
    citecolor = blue,
    filecolor = red,
    pagecolor = red,
    urlcolor = blue,
    }
\usepackage[normalem]{ulem}
\usepackage{graphicx}
\usepackage{bbold}
\usepackage{braket}
\usepackage{placeins}

\newtheorem{theorem}{Theorem}
\newtheorem{lemma}{Lemma}

\begin{document}

\title{Convex Combinations of Pauli Semigroups: Geometry, Measure and an Application}
	\author{Vinayak Jagadish}
%	\email{jagadishv@ukzn.ac.za}
	\affiliation{Quantum Research Group, School  of Chemistry and Physics,
		University of KwaZulu-Natal, Durban 4001, South Africa}\affiliation{ National
		Institute  for Theoretical  Physics  (NITheP), KwaZulu-Natal,  South
		Africa}
			\author{R. Srikanth}
%	\email{srik@poornaprajna.org}
	\affiliation{Poornaprajna Institute of Scientific Research,
		Bangalore-560 080, India}
	\author{Francesco Petruccione}
%	\email{petruccione@ukzn.ac.za}
	\affiliation{Quantum Research Group, School  of Chemistry and Physics,
		University of KwaZulu-Natal, Durban 4001, South Africa}\affiliation{ National
		Institute  for Theoretical  Physics  (NITheP), KwaZulu-Natal,  South
		Africa}
%\date{} 

\begin{abstract} 
{Finite-time Markovian channels, unlike their infinitesimal counterparts, do not form a convex set. As a particular instance of this observation, we consider the problem of mixing the three Pauli channels, conservatively assumed to be quantum dynamical semigroups, and fully characterize the resulting ``Pauli simplex.''  We show that neither the set of non-Markovian (completely positive indivisble) nor Markovian channels is convex in the Pauli simplex, and that the measure of non-Markovian channels is about 0.87.  All channels in the Pauli simplex are P divisible. A potential application in the context of quantum resource theory is also discussed.}
\end{abstract}
%\pacs{03.65.Yz,03.67.-a}
\maketitle  
\section{Introduction} Memory effects in open quantum systems~\cite{petruccione, breuer_colloquium:_2016} have become a potential experimental area of research in quantum information science. At the same time, the theoretical study of non-Markovian dynamics of open quantum systems continues to gain wide interest and poses newer conceptual challenges and surprises ~\cite{li_concepts_2017}. The dynamics of open quantum systems  are described by time-dependent positive trace preserving maps, referred to as dynamical maps. If the maps are completely positive (CP) as well, they are called quantum channels~\cite{sudarshan_stochastic_1961, Quanta77}. Interestingly, the question of what constitutes quantum non-Markovianity has remained elusive. Among various approaches are that based on CP divisibility \cite{rivas_entanglement_2010, hall2010}, on the distinguishability  of  states \cite{breuer_measure_2009}, and on the operationally motivated process-tensor formulation \cite{pollock2018non}. 

Recently, there has been an interest in convex combinations of quantum channels. An example of mixing two Markovian evolutions to create a non-Markovian one was reported in~\cite{chruscinski_non-markovianity_2015}. The idea in a more rudimentary form (before the divisibility and distinguishability criteria for quantum non-Markovianity were developed) appears in an earlier paper~\cite{wolf_assessing_2008}. In~\cite{breuer_mixing-induced_2018}, the counterintuitive behavior was explained in terms of correlations and the information flow between the system and environment. An example of a convex combination of two non-Markovian channels leading to a memoryless evolution was discussed in~\cite{wudarski2016}. In~\cite{megier_eternal_2017}, it was also shown that a master equation with an always negative decay rate arises from a mixture of Markovian semigroups. 
%.

Here, we generalize the idea of mixing channels to the case of three-way mixing, and discuss the convexity of sets of quantum (non-)Markovian dynamical maps. For simplicity, we restrict ourselves to the case of Pauli channels, whose geometrical structure is well understood~\cite{karolbook}. We show that for finite mixing of two channels, the resultant channel is non-Markovian throughout, whereas for the case of mixing three channels, a more complex picture emerges, which we fully characterize. We discuss the implications of our result and give a measure of the sets of Markovian and non-Markovian channels. 
Earlier papers~\cite{chruscinski_non-markovianity_2015,wudarski2016} have noted specific examples of nonconvexity of Markovian channels. However, this would leave open the question of whether such examples constitute a zero-measure set. Here, we characterize the full set of mixing Pauli channels (i.e., evolutions over finite time) and show that  the subset of non-Markovian channels is highly convex and that such examples have a finite (nonzero) measure. With this demonstration, the implication of the convexity of Markovian channels is significant for a quantum resource theoretic approach to non-Markovianity, which is our main result. 

Briefly, the framework for a quantum resource theory (QRT) is as follows. A QRT involves classifying all quantum states into two groups, one consisting of resource states and the other of free states \cite{chitambar2019quantum}. Associated with the free states is a set of free quantum operations that are naturally motivated and such that the set of free states is closed under the action of free operations. Further, there must be a measure of resourcefulness that must be contractive (nonincreasing) under the free operations. A QRT can address the question of what information processing tasks are (im)possible using free operations \cite{chitambar2019quantum}. In a QRT of entanglement,  the resource states are the entangled states, the free states are the separable states, while free operations are conveniently stochastic local operations and classical communication.  

In the context of studying the properties of sets of quantum channels, the Choi-Jamio\l{}kowski isomorphism \cite{Jamiolkowski-1972} can be conveniently used to represent channels acting on $d$-dimensional quantum systems as states of $d^2$-dimensional systems. This has been used to show that in contrast to the case of quantum channels, the set of \textit{generators} of CP-divisible evolutions (or equivalently, that Markovian evolutions over small-time intervals) indeed forms a convex set. This forms a basis for a convex QRT for non-Markovianity, with resource objects effectively identified with generators of CP-indivisible channels \cite{bhattacharya_convex_2018,anand_quantifying_2019}. Interestingly, \cite{berk2019resource} surveys a variety of potential QRTs for quantum non-Markovianity based on various identifications of free operations, with Markovianity defined according to the process tensor approach \cite{pollock2018non}, and identifies one of them as being the appropriate one for quantum non-Markovianity.

In our framework for a QRT of quantum non-Markovianity, the CP-divisible channels correspond to the free states and the  CP-indivisible channels to the resource states. The issue then is, what would be the appropriate free operations in this context, and is studied here. One might consider compositions of channels as a candidate for free operations. It is not hard to show that CP-divisible channels are closed under composition, denoted $\circ$. 

However, the operation $\circ$ as a free operation acting on CP indivisibility is not well defined. CP-indivisible channels can lead to entanglement with an environment, and hence their compositions will generally not lead to quantum channels. 
In this light, a natural candidate for free operations on the free states representing channels can be taken to be the action of forming convex combinations (i.e., mixing). Quantum channels are closed under mixing: given CP maps $\mathcal{E}_1$ and $\mathcal{E}_2$ mixed with fractions $\alpha$ and $(1-\alpha)$, the state $(\alpha\mathcal{E}_1 +(1-\alpha)\mathcal{E}_2)(\rho) = \alpha\mathcal{E}_1(\rho) +(1-\alpha)\mathcal{E}_2(\rho)$, which is clearly a valid state throughout the domain of validity of $\mathcal{E}_1$ and $\mathcal{E}_2$. All the same, the above results on  nonconvexity suggest that a QRT for non-Markovianity identified with CP divisibility and free operations with convex combinations would not be possible. 

In a convex QRT, the free states form a convex set. One may consider nonconvex QRTs corresponding to situations where the mixing is not available due to physical restrictions or such mixing is easily implementable and leads to resource states. A pertinent example here would be a QRT of non-Gaussianity, in the area of continuous-variable quantum information theory, where free operations taken to be Gaussian operations with adaptive feed-forward, can produce a convex combination of Gaussian states, which is in general non-Gaussian \cite{chitambar2019quantum}. 

Alternatively, one might enlarge the set of free states to the convex hull of Gaussian states, which leads to a convex QRT of non-Gaussianity, where states outside the hull are considered as ``genuinely non-Gaussian'' \cite{takagi2018convex}.
In the present context, one might ask whether a suitable enlargement of the set of free states can lead to a convex QRT of non-Markovian channels, or whether a suitable restriction on free operations would lead to a non-convex QRT for non-Markovian channels.

In this light, one might consider whether other free operations may be considered to obtain a \textit{nonconvex} QRT of CP indivisibility, such as local operations on the system, but their physical motivation is not clear. In the context of using convex combinations as free operations, one could still recover a QRT in this scenario, provided all the cases of nonconvexity could be ``quarantined'' into a zero-measure set. Thus, the main technical part of our paper is devoted to studying the geometric structure of the ``Pauli simplex,'' obtained by mixing the dynamical semigroup realizations of the three Pauli channels. This will be seen to be a simple yet nontrivial instance suitable to answer the above question.

\section{Convex Combination of Two Channels}
Consider the channel $\mathcal{E}$ acting  on a
  qubit, represented by the density matrix
\begin{equation}
  \label{eq:state}
  \rho = \frac{1}{2}(\mathbbm{1} + a_i \sigma_i) = \frac{1}{2} \left( \begin{array}{cc}1 + a_3 & a_1 - \imath a_2\\ a_1 + \imath a_2  & 1-a_3 \end{array} \right).
\end{equation}
The vector ${\bf  a} = (a_1 \,,\,  a_2 \,,\, a_3)$,
  with $|{\bf a}| \leq 1$, is the Bloch vector. Here, we consider Pauli channels which are unital, as defined by
$\mathcal{E}(\sigma_I) = \sigma_I,$ and
$\mathcal{E}(\sigma_{i}) = x_{i} \sigma_{i},
$
where $\sigma_I = \mathbbm{1}$ and $\sigma_i$'s are the Pauli matrices.

Consider the two Pauli channels
$
 \mathcal{E}_y^p(\rho)=(1-p)\rho + p\sigma_y\rho \sigma_y$ and
$\mathcal{E}_z^p(\rho) = (1-p)\rho + p\sigma_z\rho \sigma_z,$
 where $p$ is the decoherence parameter, where $p = p(t)$ is any monotonically increasing function such that $p(0)=0$ and $p(\infty)=\frac{1}{2}$. Conservatively, we choose $p$ to be $[1-\mathrm{exp}(-rt)]/2$, where $r$ is a constant. This corresponds to a quantum dynamical semigroup (QDS), with a time-independent Lindblad generator. This choice of $p$ ensuring a QDS is taken since a QDS is arguably the ``most Markovian" according to many criteria, and would thus serve to emphasize the non-convexity result that we obtain. Note that we use the same $p$ for each of the Pauli channels, making our study and analysis simple and interesting.

 \begin{theorem} 
	\label{twomixconv}
	Any finite degree of mixing of two Markovian (QDS) Pauli channels results in non-Markovianity.
\end{theorem}
For the case of $a=1$, the channel is Markovian corresponding to the two positive eigenvalues $1\pm \nu$ where $\nu= \frac{1-2q}{1-2p}$.  Theorem \ref{twomixconv} can be proven by an argument based on generators, as discussed below.

\textbf{Proof.}
Let us consider the convex combination of two channels as
\begin{eqnarray}
\mathcal{E}_\ast(p) = a\mathcal{E}_z^p + (1-a)\mathcal{E}_y^p.
\label{twomixeq}
\end{eqnarray}
The time-local generator $L$ of a channel $\mathcal{E}$, is defined by $\dot{\mathcal{E}} = L\mathcal{E}$. Quite generally (barring the existence of non-invertibility), maps and generators for the channel are equivalent and interconvertible \cite{hall2010}.
 
The differential form of the channel can be evaluated to be 
\begin{equation}
\label{megen}
L(\rho) = \sum_{k = X,Y,Z}\gamma_k(\sigma_k\rho\sigma_k-\rho),
\end{equation}
where $\gamma_k$'s are the decay rates. Note that we work in the Pauli basis $\{\mathbbm{1}, \sigma_i\}$.
The decay rate, $\gamma_X$ turns out to be
\begin{equation}
\gamma_X = -\left[\frac{ (1-a) a (1-p) p}{(1-2p) [1-2 (1-a) p] (1-2 a p)}\right]  \dot{p}.
\label{ratesdecay}
\end{equation}
For any choice of $a \in (0,1)$, it is clear that $\gamma_X$ is always negative, since expression with the square brackets and $\dot{p}$ in the right-hand side of this equation are positive. This implies that the mixing of any two Markovian Pauli channels produces a channel which is non-Markovian. 
A similar result follows for the other convex combinations of any other pair of Pauli channels as well.
\hfill $\blacksquare$

\section{Convex Combination of Three Channels}
Here, we consider the simplex obtained by arbitrary convex combinations of the three Pauli channels, which are assumed to have a QDS form and the same decay rate $c$. A general three-way mixture is described by
 \begin{equation} \tilde{\mathcal{E}}_\ast(p) = a\mathcal{E}_x^p + b\mathcal{E}_y^p+ c\mathcal{E}_z^p,
 \label{threechanneleq}
 \end{equation}
 with $a+b+c=1$ and
 $
 \mathcal{E}_x^p(\rho)=(1-p)\rho + p\sigma_x\rho \sigma_x.
$
We shall call this the \textit{Pauli simplex}. This is an equilateral triangle, whose vertices are QDS Pauli channels having a constant decay rate. 

The differential form of the channel follows to be of the same form as in Eq. (\ref{megen}), with the decay rates being
\begin{eqnarray}
\gamma_X &=& \left(\frac{1-b}{1-2 (1-b)p}+\frac{1-c}{1-2 (1-c)p}-\frac{1-a}{1-2 (1-a) p}\right)\frac{\dot{p}}{2}\nonumber\\
\gamma_Y &=&\left(\frac{1-a}{1-2 (1-a)p}+\frac{1-c}{1-2 (1-c)p}-\frac{1-b}{1-2 (1-b) p}\right)\frac{\dot{p}}{2}\nonumber\\
\gamma_Z &=&\left(\frac{1-a}{1-2 (1-a)p}+\frac{1-b}{1-2 (1-b)p}-\frac{1-c}{1-2 (1-c) p}\right)\frac{\dot{p}}{2}. \nonumber \\
\label{ratesdecaythree}
\end{eqnarray}
For arbitrary choices of $a,$ $b$ and $c$, it can be seen that in general one or more of the decay rates can become negative, implying the CP indivisibility (and thus non-Markovian nature) of the channel. The following result holds.

\begin{theorem} 
	\label{threemixconv}
	In the Pauli simplex, (a) neither the set of Markovian channels nor that of non-Markovian (CP-indivisble) maps is convex; (b) all channels are P divisible, i.e., Markovian according to the distinguishability criterion.
\end{theorem}
{\bf Proof.} The proof of (a) follows readily from Fig. \ref{fig:pauli}, derived in the following section. Here, the Markovian set is seen to constitute a curved-edge (horn) triangle within the Pauli simplex and having its vertex angles of $0^{\degree}$. To prove (b), we note that in Eq. (\ref{ratesdecaythree}), the decay rate expressions have the form 
\begin{eqnarray}
\gamma_X(a,b,c) &=& -f(a,p)+f(b,p)+f(c,p) \nonumber \\
\gamma_Y(a,b,c)  &=& f(a,p)-f(b,p)+f(c,p) \nonumber \\
\gamma_Z(a,b,c)  &=& f(a,p)+f(b,p)-f(c,p),
\label{eq:3form}
\end{eqnarray} 
where $f(\alpha,p)\ge0$ for all $p \in [0,\frac{1}{2})$ and $\alpha \in \{a,b,c\}$. The the sum $\gamma_{i} + \gamma_{j},$  $i,$ $j = X,Y, Z,$ $i \neq j$ is always positive, even though an individual rate may be negative. For example, $\gamma_X + \gamma_Y = 2f(c,p) \ge 0$.  This implies that the dynamics obtained by the mixing is P divisible and in the qubit context, is equivalent to Markovianity according to the Breuer-Laine-Piilo distinguishability criterion \cite{chruscinski_non-markovianity_2015}. \hfill $\blacksquare$

\section{Measure of (non-)Markovian maps in the Pauli triangle}
The inherent three-way symmetry in the problem helps simplify the analysis.
\begin{lemma}
	If a channel obtained as a mixture of the three Pauli (QDS) channels is non-Markovian, then precisely one of the three rates $\gamma_j$ is negative.
	\label{lem:only1}
\end{lemma}
\noindent
\textbf{Proof.} It follows from Eq. (\ref{eq:3form}) at most one of the three decay rates can be negative. Next, note that $\frac{df(\alpha,p)}{dp} = \frac{2 (1-\alpha )^2}{(1-2 (1-\alpha ) p)^2} > 0$ for all $p, \alpha$. Thus, in a given rate $\gamma_j$ ($j \in \{X, Y, Z\}$), two of the terms will produce a monotonic increase in rate, whereas the negative term will produce a monotonic decrease. Further, all rate components $\gamma_j$ start at a positive value. For example, $\gamma_Y(a,b,1-a-b,p=0) = 2b$.

Now, suppose $\gamma_Y$ is negative in some region, it follows from the considerations of the preceding paragraph that there is a $p_0$ such that $\gamma_X$ is negative for  $p \ge p_0$ and positive otherwise. That is, for $p > p_0$, we have $f(b,p) > f(a,p) + f(c,p)$, and it remains so for $p \in [p_0,\frac{1}{2})$. But, by that token, notice that $\gamma_Y$ and $\gamma_Z$ will always be positive throughout.
$\hfill \blacksquare$

\bigskip

\begin{figure}
	\includegraphics[width=8cm]{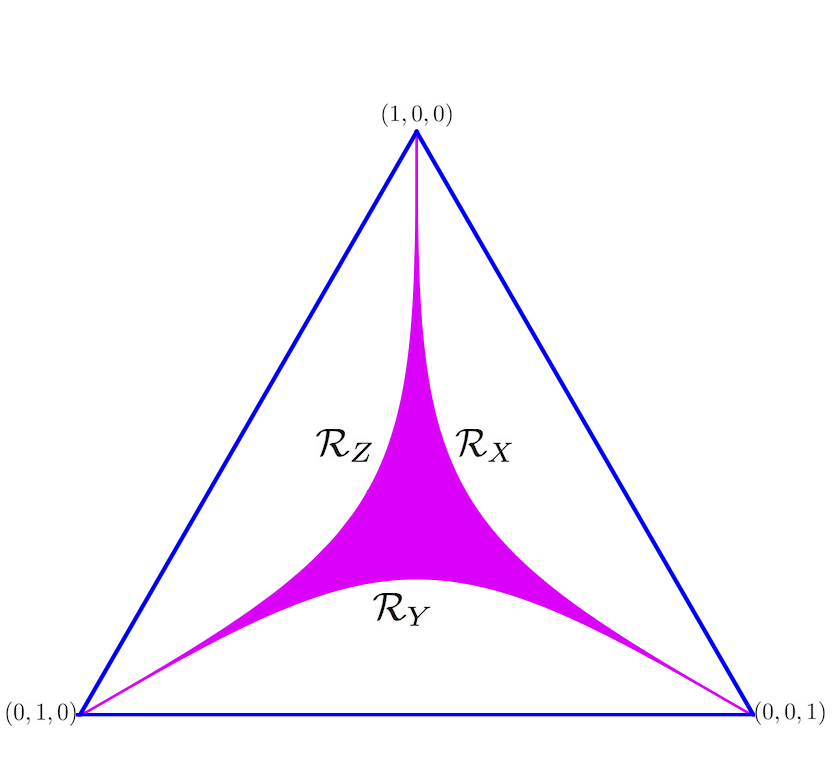}
	\caption{(Color online) The Pauli simplex (outer triangle), whose vertices are the Pauli QDS channels, and represented in convex coordinates in the Pauli-neutral representation (see text). The curved-edge (horn) triangle, with its interior colored magenta, represents the Markovian region, tapering to 0$^\circ$ at each vertex. The three convex regions marked $\mathcal{R}_j$ $(j \in \{X, Y, Z\})$ correspond to non-Markovian (CP indivisible).The corresponding edges of the horn triangle are described by the equations $\gamma_X(p=\frac{1}{2})=0$, $\gamma_Y(p=\frac{1}{2})=0,$ and $\gamma_Z(p=\frac{1}{2})=0$. The area of the horn triangle is about 0.87 of the Pauli simplex. }
	\label{fig:pauli}
\end{figure}

It follows from Lemma \ref{lem:only1} that the regions $\mathcal{R}_X$, $\mathcal{R}_Y,$ and $\mathcal{R}_Z$ in the $(a,b,c)$ parameter space, where $\gamma_X$, $\gamma_Y,$ and $\gamma_Z$ turn negative, will be non-overlapping. In Fig. \ref{fig:pauli}, the (convex) regions of negative rates are indicated by the corresponding symbol.

\begin{theorem}
The measure of non-Markovian maps in the Pauli simplex is about 0.867.
	\label{th:measure}
\end{theorem}

\noindent
\textbf{Proof.}
As noted, by virtue of the monotony of $f(\alpha,p)$, if a given rate (say) $\gamma_Y(a,b,c,p)$ turns negative at $p=p_0$, then it remains negative throughout the remaining range of $p$, and in particular at $p=\frac{1}{2}$. 

Thus, $\gamma_Y(a,b,p)$ will turn negative if and only if
\begin{eqnarray}
\gamma_Y\left(a,b,\frac{1}{2}\right) &=& \frac{a+b}{1-a-b}+\frac{1-a}{a}-\frac{1-b}{b} \nonumber \\
&\equiv& \Gamma_Y(a,b)< 0.\label{eq:step1}
\end{eqnarray}
We wish to determine the set of all points $(a,b)$ that yield negative $\gamma_Y$ at $p=\frac{1}{2}$. To this end, we solve $\gamma_Y(a,b,\frac{1}{2}-x) = 0$ for $a$ in terms of $b$, which yields
\begin{widetext}
\begin{equation}
a_\pm^Y(b,x) = \frac{1}{2} \left(\pm 
\frac{
	\sqrt{
		(b-\frac{-2 x-1}{2 x-1})
		(b - \frac{2 x+1}{2 x-1})
		(b - \frac{2-\sqrt{4 x^2-4 x+5}}{2 x-1})
		(b - \frac{\sqrt{4 x^2-4 x+5}+2}{2 x-1})
	}}
{4 (b-1) x^2-4 b x+b+1}-b+1\right).
\label{eq:aX1}
\end{equation}
\end{widetext}
From this, one finds that for $b \in [0,\beta(x)]$, where $\beta(x) \equiv \frac{2-\sqrt{4 x^2-4 x+5}}{2 x-1}$ is the region for which $\gamma_Y(a,b,0.5-x) = 0$. In this range, points $(a,b)$ such that $a \in (a_-^Y(b,x), a_+^Y(b,x))$ (respectively, $[0,a_-^Y(b,x)] \cup [a_+^Y(b,x)),1]$) represent those for which $\gamma_Y(a,b,0.5-x)$ is negative (respectively, positive). Regions of $b > \beta(x)$ are those for which $\gamma_X(a,b)$ is still positive for $p = 0.5-x$.  Thus, we fully determine $\mathcal{R}_Y$, by setting $x := 0$. Accordingly,
\begin{eqnarray}
|\mathcal{R}_Y| &=& 2  \int_{b=0}^{\beta(0)}  [a_+(b,0)-a_-(b,0)]db \nonumber\\
&=& 2  \int_{b=0}^{-2+\sqrt{5}} \frac{\sqrt{b^4+4 b^3-2 b^2-4 b+1}}{b+1} db \nonumber\\
&\approx& 0.2898,
\label{eq:mu1}
\end{eqnarray}
where the pre-factor comes from the normalization $\int_{a=0}^1 \int_{b=0}^{1-a} da ~db = \frac{1}{2}$.

Since the three non-Markovian regions are nonoverlapping, the measure of all non-Markovian channels in the Pauli simplex is 3$|\mathcal{R}_Y|$ which is 0.867. Thus, the measure of Markovian channels in the Pauli triangle is about 0.133.
\hfill $\blacksquare$ 
\bigskip

Eq. (\ref{eq:aX1}) is the equation of $\gamma_Y$ in the representation where the coordinates $(a,b)$ are used. The diagrammatic depiction of the Pauli simplex in the $(a,b)$ representation, and analogously in the $(a,c)$ or $(b,c)$ representation, is a right angle triangle. With a suitable linear transformation, one can shift to the ``Pauli neutral'' representation, which depicts the Pauli simplex as an equilateral triangle, as shown in  Fig. \ref{fig:pauli}.

The curved-edge (horn) triangle in Fig. \ref{fig:pauli}, colored in magenta, encloses the nonconvex Markovian region. The vertex angles for this triangle are all $0^{\degree}$. The convex region $\mathcal{R}_Y$, of area about 28.98\% that of the Pauli simplex, represents the set of maps where $\gamma_Y$ alone turns negative after a sufficiently long time; similarly for $\mathcal{R}_X$ and $\mathcal{R}_Z$. However, the union of these three non-Markovian regions is clearly not convex. 
Also, while the Markovian region is a connected region, the non-Markovian region is not, being the union of three disjoint regions.  

\section{Discussion and conclusions}

The measure of the non-Markovian region in the Pauli simplex being 0.87 means that if the three channels are mixed in a random proportion, then the probability that the resulting channel will be non-Markovian is about 0.87. The measure of CP-indivisible channels in the Pauli simplex can be considered as a quantification of the nonconvexity of the set of CP-divisible channels in this example. This unequivocally answers in the negative the question posed earlier, of whether the instances of nonconvexity are confined to a zero-measure set. We point out that this strongly constrains any QRT of non-Markovianity wherein CP-divisible maps correspond to free states.

\section{Acknowledgements:}
V.J. thanks Namit Anand for useful comments on the paper. The work  of V.J.  and F.P.  is based upon  research supported  by the
South African Research  Chair Initiative of the  Department of Science
and  Technology and  National Research  Foundation.  R.S.   thanks the
Defense Research  and Development  Organization (DRDO), India  for the
support      provided       through       Project No. ERIP/ER/991015511/M/01/1692.
\vspace{-3 mm}
\appendix
\section{Intermediate Maps of Convex Combinations of Channels and non-Markovianity}
Consider the convex combination of two channels as in Eq. (\ref{twomixeq}).
To check the non-Markovianity of the channel $\mathcal{E}_\ast$ according to the Rivas-Huelga-Plenio (RHP) criterion~\cite{rivas_entanglement_2010}, we consider the intermediate map $\mathcal{E}_\ast(q,p)$ defined by
 $\mathcal{E}_\ast(q) = \mathcal{E}_\ast(q,p)\mathcal{E}_\ast(p)$, with $p \le q < \frac{1}{2}$ and $a \in (0,1)$.  For this, we use the  $A$-matrix representation of the map following~\cite{sudarshan_stochastic_1961,Quanta77}. The $A$ matrix acts on the density matrix expressed as a column vector. The $A$ matrix for the intermediate map is therefore readily obtained by $A_\ast(a,q,p) = A_\ast(a,q)A_\ast(a,p)^{-1}$. By rearranging the entries of $A_\ast(a,q,p)$, one obtains the dynamical (or Choi \cite{choi_completely_1975}) matrix
 \begin{equation}
\label{eq:Bmat}
 \mathfrak{B}_\ast(a,q,p) = \frac{1}{2}\left( \begin{array}{cccc}
1+x_3 & 0 & 0 & x_1+x_2 \\ 0 & 1-x_3 & x_1 -x_2 & 0 \\ 0 & x_1 - x_2 & 1-x_3 & 0 \\ x_1+x_2 & 0 & 0 & 1+x_3
 \end{array} \right),
\end{equation} 
with
$
 x_1 =  \frac{1 -2q}{1 -2p},
x_2 = \frac{1 -2a q}{1 - 2a p}$ and
$x_3 = \frac{2a q- 2q+1}{2a p- 2p+1}$.
For complete positivity of the dynamical matrix Eq. (\ref{eq:Bmat}), and hence the Markovianity of $\mathcal{E}_\ast$, all eigenvalues of $\mathfrak{B}_\ast(a,q,p)$ must be positive. The conditions for that can be evaluated to be 
$
\abs{1\pm x_3} \geq \abs{x_{1}\pm x_{2}}.
$
For $\mathcal{E}_\ast(p)$, the corresponding dynamical matrix is of the form of Eq. (\ref{eq:Bmat}), with 
$x_1 = 1 -2p,$ $x_2 = 1 - 2a p,$ and 
$x_3 = 2a p- 2p+1.
$
One can easily see that the map $\mathcal{E}_\ast(p)$ is CP irrespective of $a$. However, the intermediate map is not-completely positive (NCP) [indicative of a negative eigenvalue for $ \mathfrak{B}_\ast(a,q,p)$] and hence non-Markovian for all $a \in (0,1)$. For instance, consider $\mathfrak{B}_\ast(0.1,0.45,0.4)$, which has a negative eigenvalue, $\approx-0.0839$, indicative of the NCP nature of the intermediate map and hence non-Markovianity. 
Thus, any nonzero mixing of two Pauli channels produces non-Markovianity, which generalizes the result of \cite{wudarski2016} (corresponding to $a=\frac{1}{2})$. This is an alternate approach to Theorem \ref{twomixconv} making use of the RHP criterion at the level of intermediate maps.

\bibliography{NMConvex.bib}
\end{document}